\documentclass[sigconf]{acmart}
\AtBeginDocument{%
  }

\usepackage{amsmath,amssymb}
\usepackage{algorithmic}
\usepackage{graphicx}
\usepackage{textcomp}
\usepackage{xcolor}
\usepackage{listings}
\usepackage{graphicx}
\usepackage{subcaption}
\setcopyright{acmlicensed}
\copyrightyear{2018}
\acmYear{2018}
\acmDOI{XXXXXXX.XXXXXXX}
\acmConference[Conference acronym 'XX]{Make sure to enter the correct
  conference title from your rights confirmation email}{June 03--05,
  2018}{Woodstock, NY}
\acmISBN{978-1-4503-XXXX-X/2018/06}

\begin{document}

\title{MulCovFuzz: A Multi-Component Coverage-Guided Greybox Fuzzer for 5G Protocol Testing}





\author{Yu Wang}
\email{hf98403@gmail.com}
\affiliation{%
  \institution{Swinburne University of Technology}
  \country{Australia}
}

\author{Yang Xiang}
\email{yxiang@swin.edu.au}
\affiliation{%
  \institution{Swinburne University of Technology}
  \country{Australia}
}

\author{Chandra Thapa}
\email{chandra.thapa@data61.csiro.au}
\affiliation{%
  \institution{CSIRO's Data61}
  \country{Australia}
}

\author{Hajime Suzuki}
\email{Hajime.Suzuki@data61.csiro.au}
\affiliation{%
  \institution{CSIRO Data61}
  \country{Australia}
}


\begin{abstract}
As mobile networks transition to 5G infrastructure, ensuring robust security becomes more important due to the complex architecture and expanded attack surface. Traditional security testing approaches for 5G networks rely on black-box fuzzing techniques, which are limited by their inability to observe internal program state and coverage information. This paper presents MulCovFuzz, a novel coverage-guided greybox fuzzing tool for 5G network testing.
Unlike existing tools that depend solely on system response, MulCovFuzz implements a multi-component coverage collection mechanism that dynamically monitors code coverage across different components of the 5G system architecture. Our approach introduces a novel testing paradigm that includes a scoring function combining coverage rewards with efficiency metrics to guide test case generation.
We evaluate MulCovFuzz on open-source 5G implementation OpenAirInterface. Our experimental results demonstrate that MulCovFuzz significantly outperforms traditional fuzzing approaches, achieving a 5.85\% increase in branch coverage, 7.17\% increase in line coverage, and 16\% improvement in unique crash discovery during 24h fuzzing testing. MulCovFuzz uncovered three zero-day vulnerabilities, two of which were not identified by any other fuzzing technique. This work contributes to the advancement of security testing tools for next-generation mobile networks.
\end{abstract}



\received{20 February 2007}
\received[revised]{12 March 2009}
\received[accepted]{5 June 2009}

\maketitle

\section{Introduction}

Finding vulnerabilities in 5G networks is critically important as these networks serve as the backbone of modern digital infrastructure, connecting billions of users worldwide and enabling transformative technologies \cite{agiwal2016next}. With over 5 billion unique mobile subscribers and 12 billion mobile connections as of 2024, security flaws could have massive widespread impact, potentially affecting millions of devices per square kilometer due to 5G's high connection density \cite{agiwal2016next, chettri2019comprehensive}. The complexity of 3GPP specifications and implementation in memory-unsafe languages like C/C++ increases the likelihood of vulnerabilities, while core protocols like PFCP could be exploited for attacks ranging from denial of service to traffic hijacking and billing fraud \cite{amponis2022threatening}. As 5G increasingly underlies critical infrastructure, IoT networks, edge computing, and cloud-native applications \cite{gupta2015survey, li20185g}, security breaches could result in severe service disruptions, financial losses, and compromise of sensitive data. Therefore, proactive vulnerability discovery through techniques like fuzzing and security testing is essential to ensure 5G networks remain robust and trustworthy as they become more central to modern society \cite{rost2016mobile}.

5G mobile networks introduce several revolutionary technologies compared to previous generations, including Software-Defined Networking (SDN) and Network Function Virtualization (NFV) that enable network control to be decoupled from hardware and allow network functions to be virtualized \cite{lei20215g}; Network Slicing that allows creating multiple virtual networks over shared infrastructure for service differentiation \cite{network20113rd}; Multi-access Edge Computing (MEC) that brings computing resources closer to users to reduce latency \cite{network20113rd}; Service-Based Architecture (SBA) with modular network functions exposed as services \cite{rommer20195g}; cloud-native design with microservices and containerization for better resource utilization \cite{shah2021cloud, dao2021survey}; and Open and Programmable RAN with disaggregation of RAN components and open interfaces between them \cite{thantharate2019deepslice}. Together, these technologies enable 5G networks to support enhanced mobile broadband (eMBB), ultra-reliable low latency communications (URLLC), and massive machine type communications (mMTC) use cases with unprecedented flexibility, efficiency and performance compared to previous generations \cite{shah2021cloud}.

These new 5G technologies introduce significant testing challenges from both functional and non-functional perspectives. From a functional perspective, the disaggregation of network components and introduction of new interfaces require comprehensive interoperability and integration testing \cite{salazar20215greplay}. The non-functional testing challenges include validating performance metrics like throughput, latency, and scalability across virtualized network functions and slices \cite{shah2021cloud}. Additionally, the Service-Based Architecture (SBA) of 5G introduces new security vulnerabilities, as traditional security and privacy mechanisms become ineffective or inapplicable in the 5G context \cite{hu2022fuzzing}. This necessitates the development of specialized security test cases and tools specifically designed for 5G networks \cite{potnuru2021berserker}. The dynamic nature of network slicing and edge computing also demands new approaches to test the robustness and reliability of resource allocation and service delivery \cite{thantharate2019deepslice}. The ENISA threat landscape report highlights how AMFs are vulnerable to replay attacks of Non-Access Stratum (NAS) signaling messages between the UE and AMF on the N1 interface \cite{paskauskas2023enisa}. Similarly, malicious actors can execute replay attacks by manipulating Packet Forwarding Control Protocol (PFCP) messages that manage GPRS Tunnelling Protocol (GTP) tunnels \cite{ahmad2018overview}. These vulnerabilities could lead to serious consequences including denial of service, information leakage, and network instability. A practical example comes from recent testing of the free5GC implementation, where replayed NAS Security Mode Complete messages successfully triggered security context anomalies in the AMF, highlighting the need for robust testing mechanisms \cite{salazar20215greplay}.

Traditional security testing approaches for 5G networks have primarily relied on stateful black-box fuzzing techniques, which are limited by their inability to observe internal program states and coverage information \cite{garbelini2022towards}. While these approaches have identified various vulnerabilities, they fail to leverage recent advances in coverage-guided greybox fuzzing (CGF) that have demonstrated significant improvements in bug-finding capabilities across other domains \cite{bohme2016coverage, fioraldi2020afl++, hu2024enhancing}. To address these limitations, we propose a comprehensive fuzzing framework that incorporates coverage feedback to guide the testing process. Unlike existing 5G fuzzing tools that depend solely on server under test (SUT) response feedback \cite{fioraldi2020afl++, salazar20215greplay} and traditional grey-box fuzzing tool only rely on single coverage information \cite{pham2019smart, yu2020sgpfuzzer, pham2020aflnet}, our method implements a multi-component coverage mechanism and uses it to guide test case generation through a feedback-driven approach. Given the intricate interdependencies among components in 5G system architecture, MulCovFuzz applies instrumentation to each individual component and gathers their corresponding coverage information. Leveraging this information, the framework calculates an energy score for each test seed, guiding the allocation of fuzzing time such that seeds with higher scores receive proportionally more testing resources. Through this systematic instrumentation approach, we can track how fuzz-generated packets propagate through the entire 5G infrastructure and analyze their effects. Rather than limiting our analysis to localized component behaviors, our methodology enables us to observe ripple effects across components and capture complex interaction patterns that emerge from their interdependencies.

In summary, our paper makes the following contributions:

\begin{itemize}

    \item We introduce a novel coverage-guided greybox fuzzing tool for 5G core network system. Unlike traditional approaches that rely solely on system responses, our framework dynamically adapts its fuzzing strategies based on code coverage information, enabling more thorough exploration of protocol state spaces and critical execution paths.
    \item We develop a multi-component coverage collection mechanism that calculate coverage separately from different components in the 5G system. Utilizing the multi-component coverage collection mechanism, it is possible to more accurately estimate the impact of the packet sent by the fuzzer on the 5G system.
    \item We implement and evaluate our tool across on open-source implementation \textit{OpenAirInterface}.
    \item We design a novel scoring function for energy assignment based on the multi-component coverage.
    \item We provide a comprehensive framework for analyzing different fuzzing strategies in 5G protocols.
\end{itemize}

\section{Motivation Example}

\lstset{
  language=C++,
  breaklines=true,
  basicstyle=\footnotesize\ttfamily,
  keywordstyle=\color{blue}\bfseries, 
  commentstyle=\color{gray}\itshape,   
  stringstyle=\color{red}             
}

\begin{figure*}[tb]
\begin{lstlisting}[
    caption={Code paths in AMF that depend on interactions with other components. Traditional AMF-focused fuzzers tend to miss the error handling and corner case paths that require monitoring multi-component interactions.},
    label={lst:amf_code_paths},
    basicstyle=\footnotesize\ttfamily,
    numbers=left,
    numberstyle=\tiny,
    breaklines=true,
    frame=single,
    language=C++
]
// AMF function handling PDU Session Resource Setup
void amf_app::handle_pdu_session_resource_setup_request(
        const ngap_pdu_session_resource_setup_request_t& request) {

    // Path 1: Standard flow - Always covered by traditional fuzzers
    if (validate_request(request)) {
        process_standard_request(request);
        return;
    }

    // Path 2: Error handling - Rarely covered by traditional fuzzers
    if (smf_session_status != SMF_SESSION_ACTIVE) {
        handle_smf_session_error(request, smf_session_status);
        return;
    }

    // Path 3: Corner case - Almost never covered by traditional fuzzers
    if (nrf_discovery_result == NRF_LIMITED_DISCOVERY &&
        pdu_session_type == PDU_SESSION_TYPE_ETHERNET &&
        upf_config_status == UPF_CONFIG_PARTIAL) {
        handle_specialized_configuration(request);
        return;
    }

    // Default handling
    handle_default_case(request);
}
\end{lstlisting}
\end{figure*}

To illustrate the limitations of traditional fuzzing approaches and highlight the need for multi-component coverage feedback in 5G core network testing, we present a concrete example from the AMF's PDU Session Resource Setup handling. This example demonstrates how certain code paths in AMF become accessible only when considering its interactions with other network functions, as shown in Listing~\ref{lst:amf_code_paths}.

In this example, the AMF’s handling of the PDU Session Resource Setup request reveals several critical code paths. \textbf{Path 1 (Standard Flow)} represents the typical execution path, which traditional gray-box fuzzers—focusing solely on a single component’s code coverage—can easily reach by sending well-formed PDU Session Resource Setup requests. \textbf{Path 2 (Error Handling)} corresponds to a less common scenario triggered when the SMF reports a non-active session status. Traditional AMF-focused fuzzers, such as AMFuzz, often fail to reach this path because they lack visibility into the SMF’s internal state and cannot account for how SMF responses influence AMF’s behavior. \textbf{Path 3 (Complex Corner Case)} exemplifies an intricate interaction across multiple components (NRF, SMF, and UPF). This path is rarely exercised in traditional fuzzing campaigns because it requires a precise combination of internal states across different network functions—conditions that are nearly impossible to satisfy without coordinated multi-component feedback.

Consequently, traditional fuzzers encounter significant limitations when testing such multi-component interaction scenarios:

\begin{itemize}
    \item There is a lack of dedicated gray-box fuzzing frameworks tailored to the complexity of 5G core network environments.
    \item Traditional protocol fuzzing approaches may achieve high code coverage within the SUT but often overlook inter-component dependencies, making them difficult to apply effectively in 5G core network testing.
    \item These approaches typically emphasize shallow protocol parsing paths, while neglecting deeper functional logic that relies on interactions with other network components.
\end{itemize}

To address these limitations, we propose a \textit{Multi-Coverage Guided Fuzzing} framework, introducing a novel testing paradigm tailored for the 5G core network.

To overcome Limitations 1 and 2, we design a 5G-aware gray-box fuzzing framework along with a novel \textit{Coverage Collection Module}, which is capable of gathering distinct coverage information from each network component interacting with the AMF. 

To mitigate Limitation 3, we introduce a carefully designed scoring function that effectively guides the fuzzer toward exploring more complex and interaction-dependent behaviors within the SUT.

\section{5G System architecture and interfaces}

The 5G Core Network (5GC) represents a revolutionary advancement in mobile network architecture, introducing a cloud-native, service-based architecture (SBA) standardized by 3GPP in Release 15 \cite{3gpp}. At its core, 5GC employs a modular framework of Network Functions (NFs) that communicate through standardized service-based interfaces (SBIs) using HTTP/2 and RESTful APIs \cite{dolente2023vulnerability}. Key components include the Access and Mobility Management Function (AMF) for mobility management, Session Management Function (SMF) for session control, User Plane Function (UPF) for data routing \cite{dolente2023vulnerability} and so on. This architecture introduces groundbreaking capabilities like network slicing, allowing operators to create multiple virtual networks over single physical infrastructure to support diverse use cases \cite{dolente2023vulnerability, cardoso2020softwarized}. The 5GC also implements enhanced security mechanisms and supports both standalone and non-standalone deployments \cite{dolente2023vulnerability}, facilitating flexible migration from 4G networks while enabling comprehensive 5G capabilities. This architectural framework serves as the foundation for advanced 5G applications and future mobile network evolution, supporting technologies like network function virtualization (NFV) and software-defined networking (SDN) \cite{dolente2023vulnerability, cardoso2020softwarized}.

The 5G system architecture implements several critical interfaces for enabling communication between different network functions and components \cite{garbelini2022towards}. At the core of this architecture, the N1 interface facilitates Non-Access Stratum (NAS) signaling between User Equipment (UE) and Access and Mobility Management Function (AMF), while the N2 interface supports the Next Generation Application Protocol (NGAP) communication between the Next Generation NodeB (gNB) and AMF \cite{dolente2023vulnerability}. The N3 interface connects gNB to User Plane Function (UPF) for user plane data transmission, while N4 interface enables control plane interaction between Session Management Function (SMF) and UPF through the Packet Forwarding Control Protocol (PFCP) \cite{nakas20235g, mancini2024amfuzz}. The N6 interface links UPF to Data Network (DN), facilitating external connectivity. Within the Service-Based Architecture (SBA), Network Functions (NFs) communicate via standardized Service-Based Interfaces (SBIs), implemented using RESTful APIs over HTTP/2 protocol. These SBIs, designated as Nx interfaces (where x represents specific NF pairs), enable flexible service-oriented interactions between NFs \cite{dolente2023vulnerability}.

\section{Methodology}

\begin{figure*}
    \centering
    \includegraphics[width=1\linewidth]{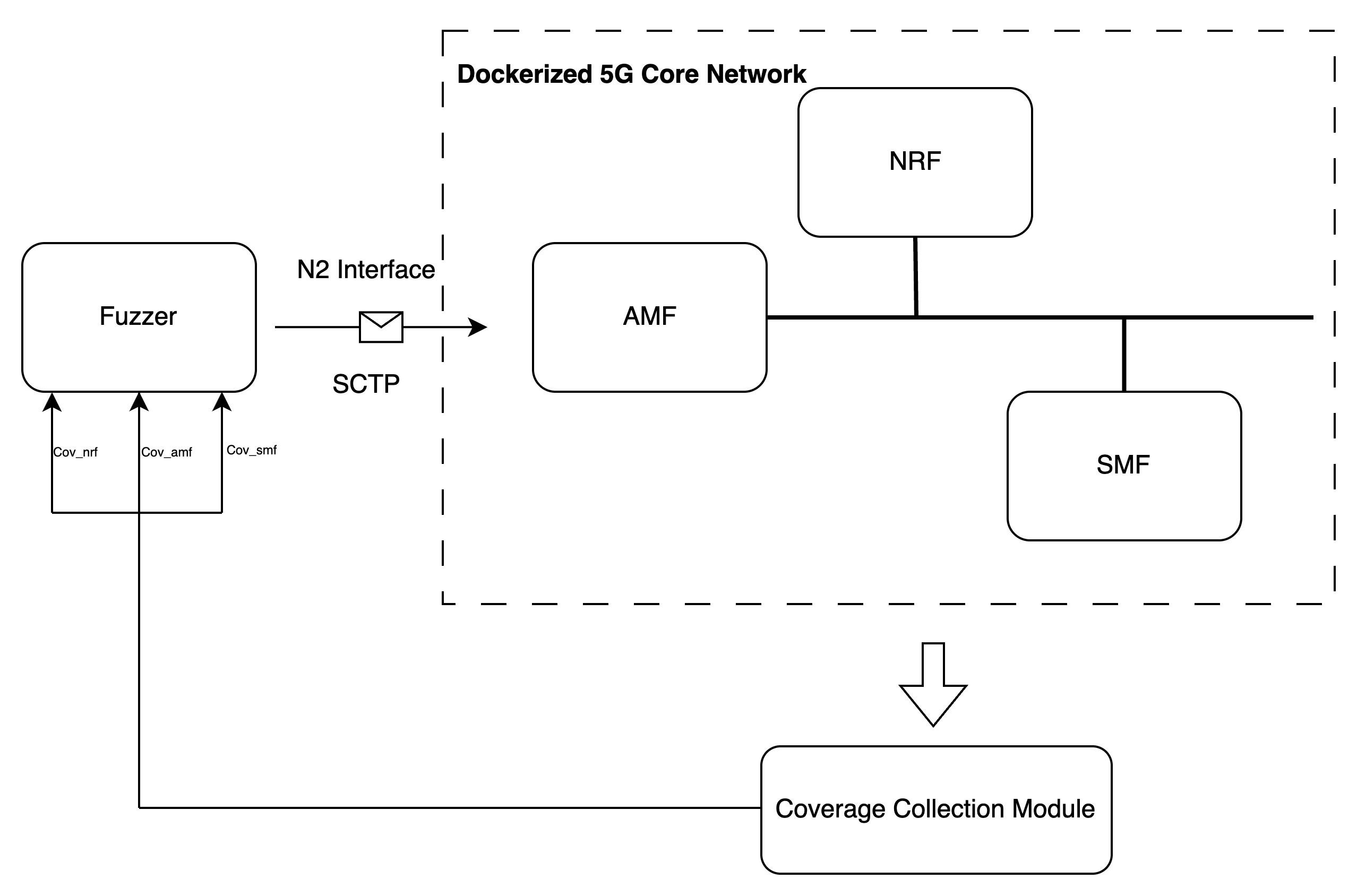}
    \caption{The overall framework of MulCovFuzz}
    \label{fig:overall_framework}
\end{figure*}

\subsection{Overview Framework}

Fig.~\ref{fig:overall_framework} illustrates the overall architecture of \textit{MulCovFuzz}. In this framework, the 5G core network is composed of multiple distributed components (e.g., AMF, SMF, NRF and UPF) that communicate with each other over IP-based interfaces. MulCovFuzz operates as a client, while the AMF serves as the main server under test in this setting.

MulCovFuzz interacts with the SUT through three types of feedbacks:

\begin{itemize}
    \item \textbf{Message Feedback:} Responsible for sending test inputs to the 5G core network and receiving responses. In particular, messages are sent to the AMF via the N2 interface over SCTP.
    
    \item \textbf{Coverage Feedback:} Used to collect code coverage feedback. The main channel is dedicated to gathering code coverage information from the Components.
    
    \item \textbf{Signal Feedback:} Monitors the runtime status of the AMF process. MulCovFuzz uses this channel to detect whether the target system experiences anomalies such as crashes or abnormal terminations during test execution.
\end{itemize}

The testing process begins when MulCovFuzz sends a test request to the AMF through the N2 interface. Upon receiving the request, the AMF processes it and sends a corresponding response back to the fuzzer. Meanwhile, the signal feedback continuously monitors for crashes or abnormal behaviors triggered by the test input.

Next, MulCovFuzz collects initial main coverage information along with AMF's response. Based on this information, it determines whether to invoke the \textit{Coverage Collection Module} to gather additional coverage data from other components.

Once all relevant coverage data is collected, MulCovFuzz calculates a global contribution score for the current seed. This score reflects how effectively the input explores multi-component behaviors. The scheduling module then prioritizes seeds with higher contributions by allocating more testing resources to them.

\subsection{Coverage Feedback Construction}

\subsubsection{Instrumentation}

Prior to conducting fuzzing activities, same as AFLNET instrument operation \cite{pham2020aflnet}, MulCovFuzz utilizes afl-clang-fast tool to instrument various components of the 5G core network based on LLVM framework to track code coverage changes during testing. Specifically, we will instrument key Network Functions (NFs) including AMF, SMF, UPF, NRF, and NEF. The instrumentation will collect basic block coverage. This systematic coverage monitoring approach enables us to analyze coverage trends, identify uncovered code regions, evaluate test case effectiveness, and guide subsequent testing strategies.

\subsubsection{Coverage construction}

For efficient coverage feedback collection, we implement a communication mechanism between the 5G system components and the fuzzer. Each component $i$ maintains a coverage bitmap $B_i$ in shared memory (SHM) that tracks edge hit counts during execution. The bitmap is structured as:

\begin{equation}
    B_i = \{b_1, b_2, ..., b_m\}, b_j \in \{0,1,2,3,4,5,6,7\}
\end{equation}

The edge coverage quantization system implements an 8-level (0-7) logarithmic bucketing scheme, following the methodology established in the American Fuzzy Lop (AFL) fuzzer. Each bucket $b_j$ corresponds to a specific range of edge hit frequencies, where $b_0$ indicates unhit edges (0 hits), $b_1$ represents exactly 1 hit, $b_2$ denotes exactly 2 hits, $b_3$ encompasses 3-4 hits, $b_4$ covers 5-8 hits, $b_5$ spans 9-16 hits, $b_6$ represents 17-32 hits, and $b_7$ indicates more than 32 hits. The parameter i indicates the number of distinct feedback channels in the system, where each channel corresponds to a unique component in the 5G architecture. MulCovFuzz aggregates coverage feedback information across various components and leverages this multi-component coverage data to guide the fuzzing process. This multi-component feedback mechanism enables MulCovFuzz to generate test cases that comprehensively exercise the entire 5G system rather than being limited to testing individual components in isolation. The integration of coverage data from multiple architectural components allows for a more holistic assessment of the system's behavior and potential vulnerabilities.

\subsection{Multi-Component Coverage Collection Module}

\begin{figure*}
    \centering
    \includegraphics[width=1\linewidth]{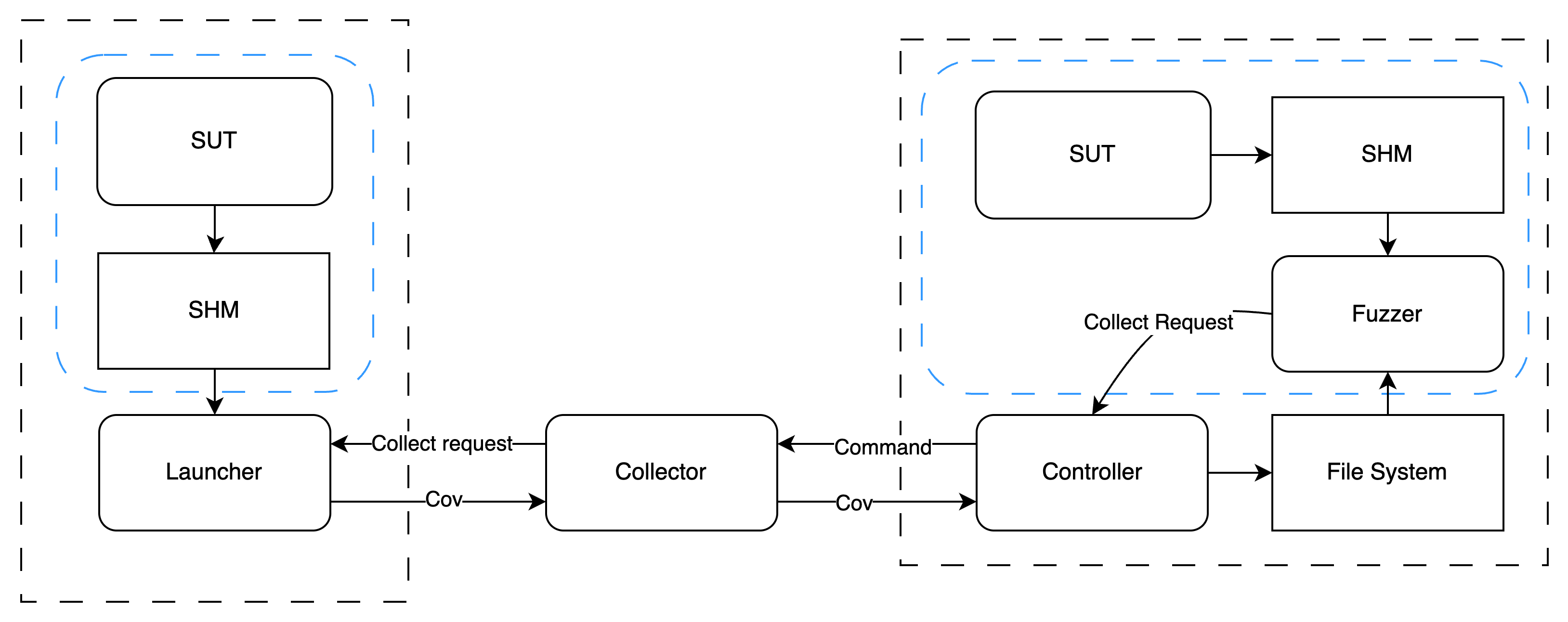}
    \caption{The workflow of Multi-Coverage Collection Module}
    \label{fig:MCCM_workflow}
\end{figure*}

Due to the distributed nature of 5G core networks and the latency introduced by inter-component communication, real-time coverage monitoring across multiple components can significantly degrade fuzzing throughput. This poses a major challenge for collecting accurate and efficient multi-component coverage feedback.

To address this issue, we designed the \textit{Multi-Coverage Collection Module (MCCM)}. As shown in Fig.~\ref{fig:MCCM_workflow}, the module is composed of three main components: \textbf{Controller}, \textbf{Collector}, and \textbf{Launcher}, each of which plays a distinct role in managing and aggregating coverage data from distributed 5G components.

\subsubsection{Controller}

The Controller is responsible for processing collection requests issued by the fuzzer and coordinating all feedback from components other than the AMF. During fuzzing, MulCovFuzz first gathers AMF's coverage from the SHM and evaluates whether the new input has triggered previously unseen paths (i.e., new bits in the coverage bitmap).

If a potentially interesting change is observed, MulCovFuzz sends a collection request to the Controller, indicating that further analysis is needed to determine whether this input also exercises new paths in other components.

Upon receiving the request, the Controller communicates with each remote Collector via lightweight Remote Procedure Call mechanisms. Specifically, it sends collection commands through socket-based communication channels, ensuring compatibility within containerized environments.

Once the Collectors retrieve coverage data from their respective containers, they serialize the results and return them to the Controller using the same communication channel. The Controller then aggregates the results and writes them into a shared I/O buffer monitored by the fuzzer process. This I/O-based feedback mechanism allows the fuzzer to seamlessly integrate multi-component coverage data into its mutation and scheduling logic.

This remote triggering and asynchronous feedback design enables efficient and scalable coordination between distributed fuzzing agents and their coverage sources in a 5G testing environment.

\subsubsection{Collector}

Each Collector is deployed within a dedicated container alongside its associated target component. Its primary role is to respond to commands from the Controller, access the local SHM region, and extract the most recent code coverage data.

Upon receiving a command, the Collector identifies the corresponding shared memory segment, which is typically created using POSIX shared memory APIs (e.g., \texttt{shm\_open}, \texttt{mmap}). Each component writes coverage information, usually in the form of a bitmap, to this region during execution.

The Collector maps this SHM segment into its address space, reads the coverage data, and performs a lightweight encoding or compression step if necessary to reduce communication overhead. The encoded coverage data is then packaged into a structured response and transmitted back to the Controller via the established socket connection.

To ensure data consistency and avoid race conditions, the Collector may use synchronization primitives such as memory barriers or file locks when reading from SHM. This guarantees that the retrieved coverage reflects the complete execution state of the component for the given test case.

This mechanism enables low-latency and accurate extraction of multi-component coverage information within the containerized 5G environment.

\subsubsection{Launcher}

The Launcher is responsible for initializing and managing the runtime environment for each edge component. It performs three main tasks as part of the setup procedure.

First, it launches the edge component within its dedicated container. This is done using container orchestration commands (e.g., via Docker CLI or Docker SDK), with the necessary runtime configurations such as network isolation and mounted volumes.

Second, the Launcher creates a SHM segment inside the container to store runtime code coverage data. This SHM region is established using POSIX SHM APIs (e.g., \texttt{shmget}, \texttt{shmat}) and is exposed to both the instrumented component and the Collector process. Each component is compiled with AFL-style instrumentation, which inserts coverage counters into the binary during build time. These counters are automatically updated during execution and reflect which code paths have been executed. The runtime coverage bitmap is then continuously maintained in the SHM region at a fixed memory address defined by AFL's instrumentation conventions (typically referenced by the \texttt{\_\_afl\_area\_ptr} pointer).

Third, the Launcher monitors the status of the component process to ensure it remains alive and responsive throughout the fuzzing cycle. This includes checking the process health via system-level probes and relaunching the component if it crashes or exits unexpectedly.

By reusing AFL-compatible instrumentation and coverage collection infrastructure within each component container, MulCovFuzz achieves seamless integration with existing greybox fuzzing workflows while enabling fine-grained, multi-component feedback with minimal performance overhead.

\subsection{Scoring function}

The scoring function evaluates test cases through a weighted combination of coverage reward and efficiency reward, expressed as

\begin{equation}
    S(t) = w_1R_c(t) + w_2R_e(t)
\end{equation}
coverage reward $R_c(t)$ quantifies a test case's contribution to code coverage across all instrumented channels using

\begin{equation}
    R_c(t) = \sum_{i=1}^{n} \alpha_i C_i(t)
\end{equation}
where $\alpha_i$ represents the weight coefficient for each channel and $C_i(t)$ measures the normalized coverage gain in that channel. The efficiency reward $R_e(t)$ assesses resource utilization through

\begin{equation}
    R_e(t) = \beta \cdot \frac{1}{T(t)} \cdot \sum_{i=1}^{n} C_i(t)
\end{equation}
where $T(t)$ represents execution time and $\beta$ serves as a scaling factor. This dual-reward mechanism guides the fuzzer to prioritize test cases that discover new paths across multiple channels while maintaining computational efficiency, with weights $w_1$, $w_2$ and channel coefficients $\alpha_i$ that can be dynamically adjusted to optimize the exploration-exploitation tradeoff.

\section{Experiment}

\subsection{Implementation and setting}

MulCovFuzz is implemented in C/C++, which extended by AFLNet. For empirical evaluation and comparison with state-of-the-art fuzzers targeting 5G systems, we utilize open-source 5G core network implementations: OpenAirInterface (OAI). These currently represent the most widely adopted open-source 5G core network projects with functional implementations. MulCovFuzz is designed to evaluate system stability by sending packets to the access network of these 5G systems. The experimental evaluation was conducted on a high-performance workstation equipped with an AMD Ryzen Threadripper 2990WX 32-Core Processor, 128GB RAM, running Ubuntu 20.04 LTS (64-bit) as the operating system.


\subsection{Coverage Analysis}

In this section, we evaluate the fuzzing effectiveness with respect to the RAN-Core interface, specifically targeting the Access and Mobility Management Function (AMF) component. Rather than analyzing the entire 5G core network stack, we focus on the AMF due to its central role in handling signaling interactions from the Radio Access Network (RAN) over the N2 interface for initial protocol message exchanges.

We instrument the AMF component using LLVM-based coverage tools to collect runtime branch and line coverage data during fuzzing. The goal is to assess how different fuzzing strategies, including a baseline and our proposed multi-component coverage guided approach, perform in exploring the AMF’s internal logic, particularly complex or rarely triggered code paths.

\subsubsection{OAI}

For the OAI 5G core implementation, we conduct a detailed coverage analysis of the AMF module, which is responsible for handling N2 signaling procedures such as Registration, PDU Session Setup, and Service Request. Our fuzzing experiments specifically target these procedures to evaluate how thoroughly each fuzzing strategy exercises the underlying logic.

\begin{figure*}[htbp]
    \centering
    \begin{subfigure}[b]{0.45\linewidth}
        \centering
        \includegraphics[width=\linewidth]{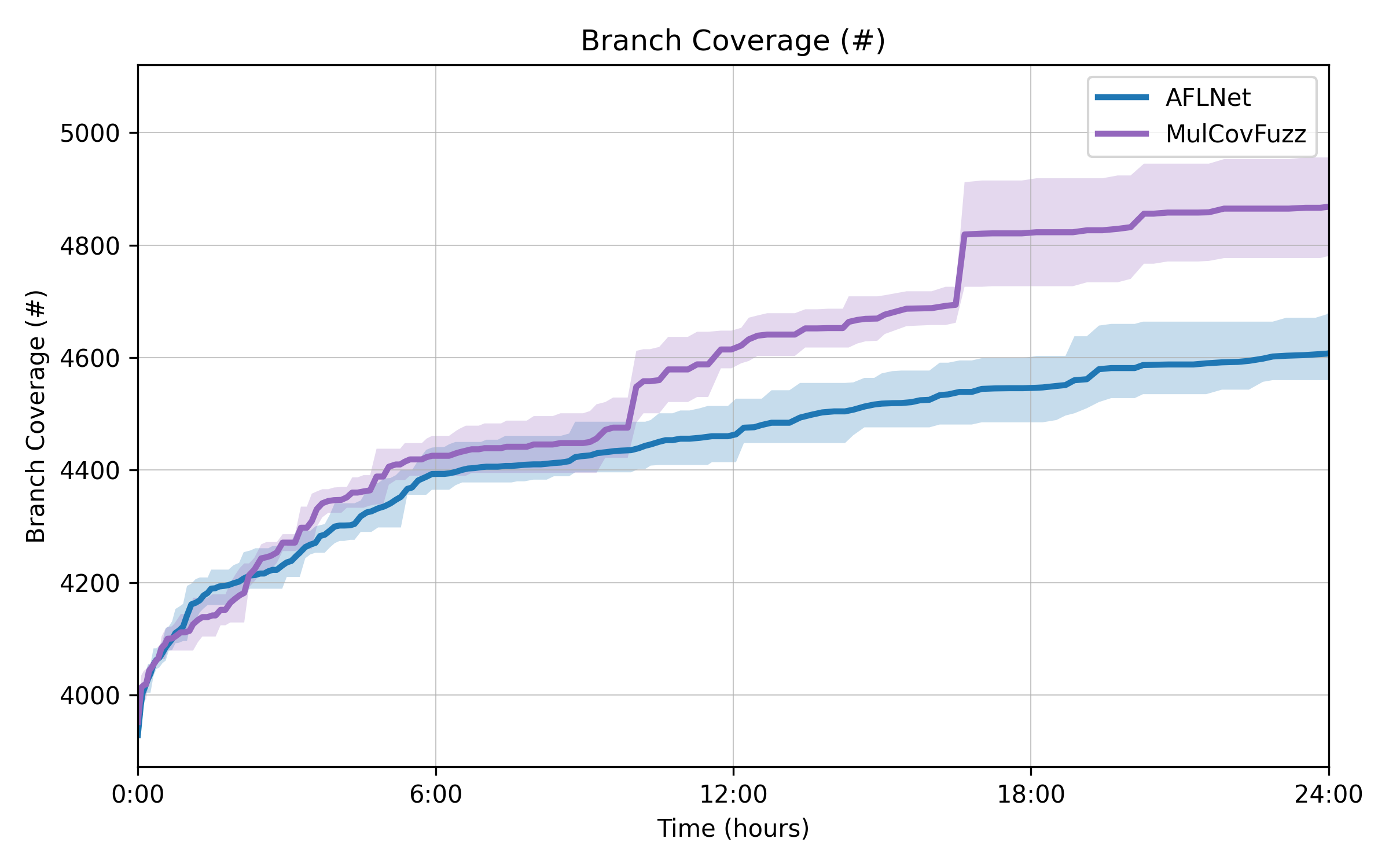}
        \caption{The number of branches covered}
        \label{fig:sub1}
    \end{subfigure}
    \hfill
    \begin{subfigure}[b]{0.45\linewidth}
        \centering
        \includegraphics[width=\linewidth]{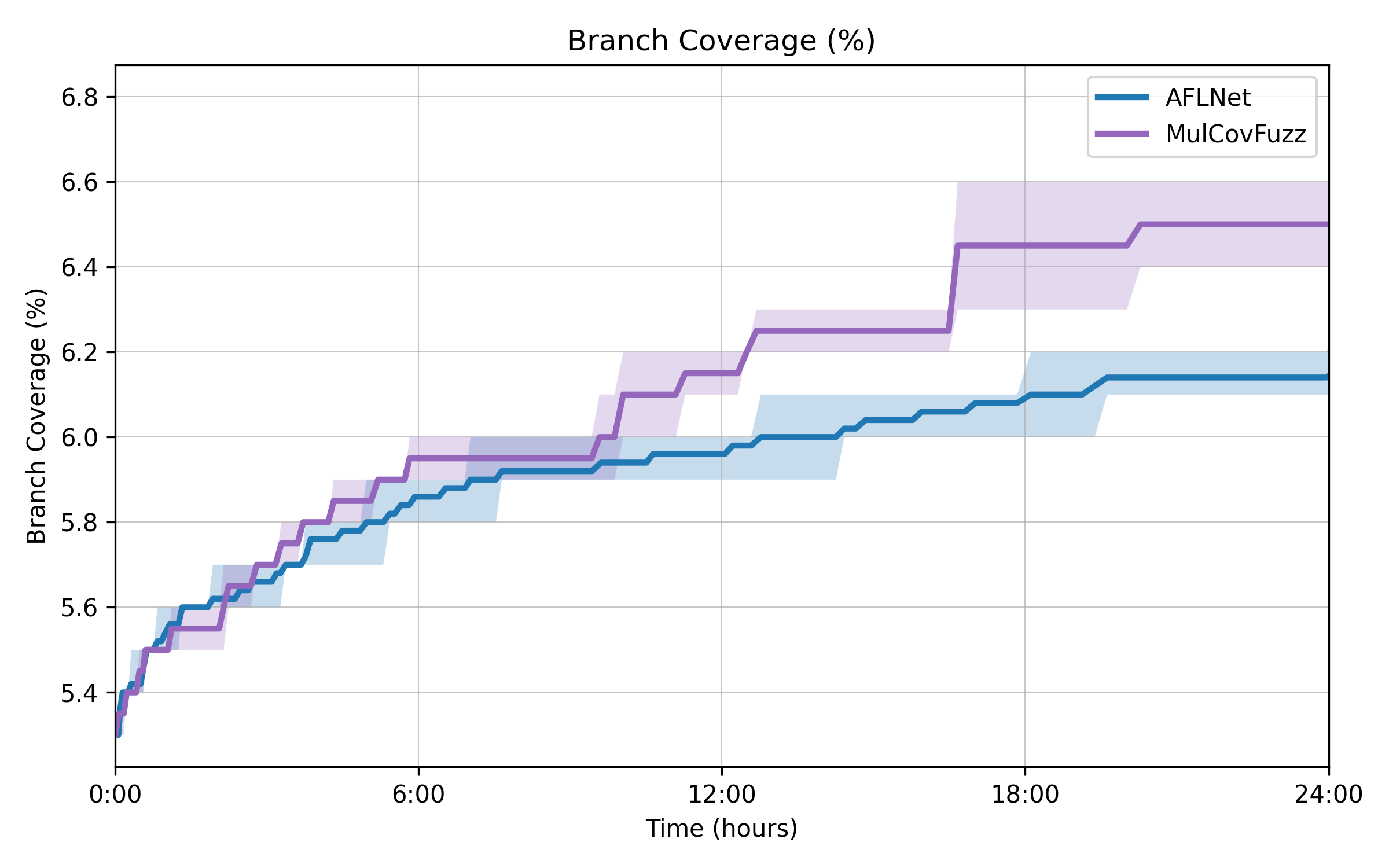}
        \caption{The proportion of branch coverage}
        \label{fig:sub2}
    \end{subfigure}
    \hfill
    \begin{subfigure}[b]{0.45\linewidth}
        \centering
        \includegraphics[width=\linewidth]{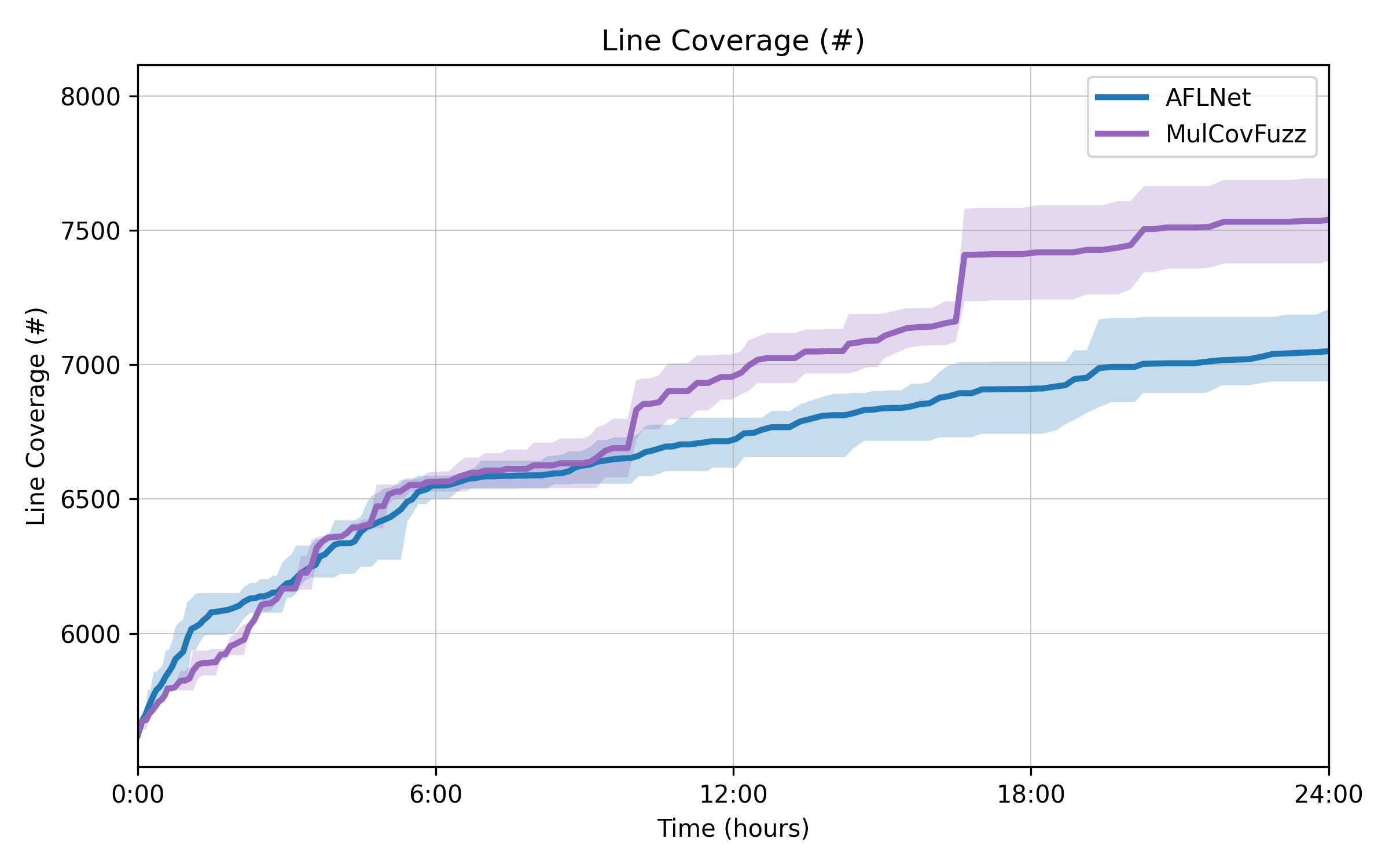}
        \caption{The number of lines covered}
        \label{fig:sub3}
    \end{subfigure}
    \hfill
    \begin{subfigure}[b]{0.45\linewidth}
        \centering
        \includegraphics[width=\linewidth]{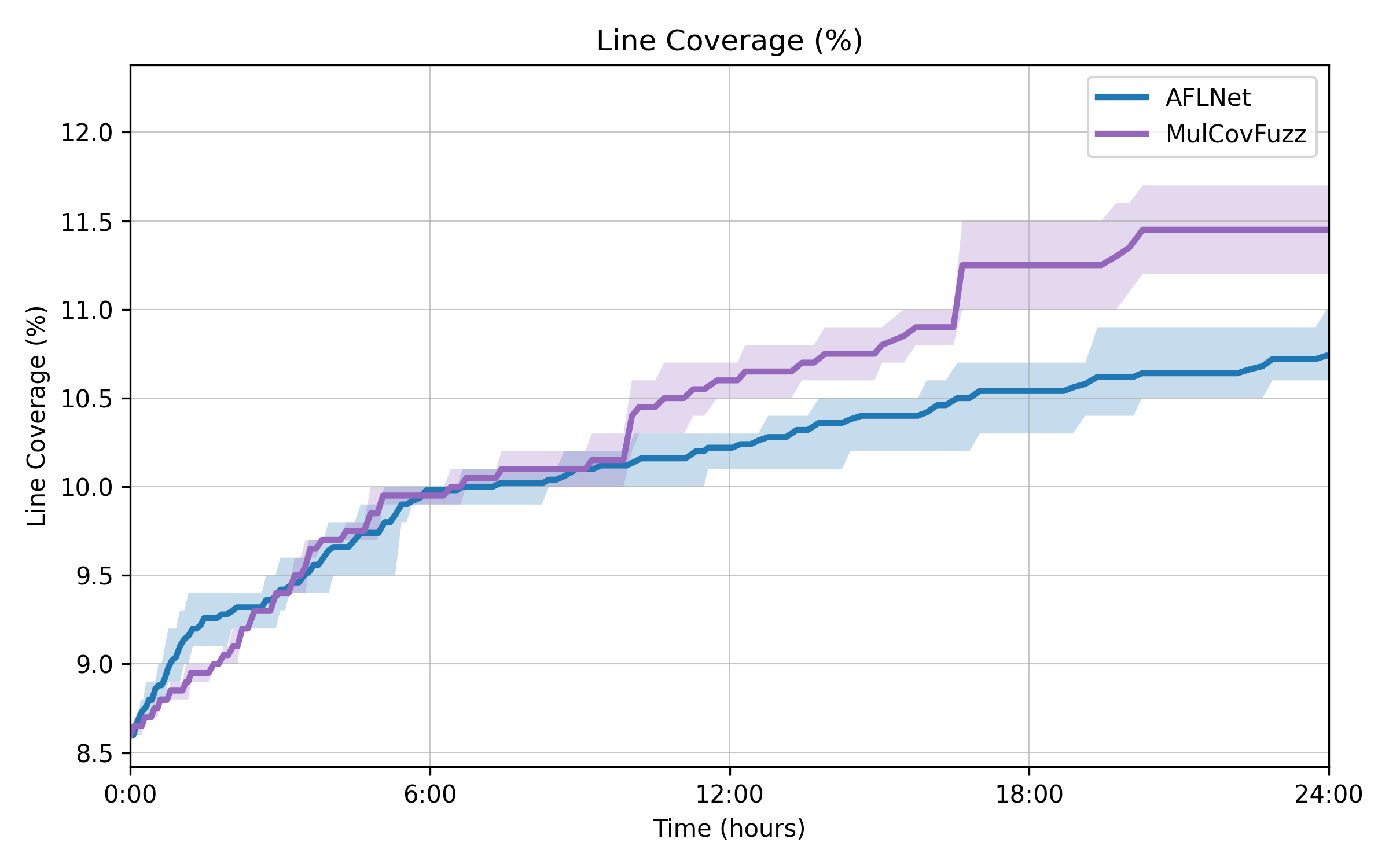}
        \caption{The proportion of line coverage}
        \label{fig:sub4}
    \end{subfigure}
    \caption{Coverage performance comparison about different fuzzing tools in OAI-AMF}
    \label{fig:cov-oai}
\end{figure*}

Each fuzzing configuration is evaluated over five independent 24-hour trials, and all coverage results are averaged accordingly. In Figure~\ref{fig:cov-oai}, the solid lines represent the mean coverage across the five runs, while the shaded regions indicate the range from minimum to maximum values observed.

Figure~\ref{fig:cov-oai} demonstrates that our proposed multi-component coverage guided fuzzing approach significantly outperforms the baseline AFLNet fuzzer across all metrics. On average, our method achieves a 5.85\% increase in branch coverage and a 7.17\% increase in line coverage, effectively uncovering deeper and more complex execution paths within the AMF logic. These results underscore the importance multiple coverage signals to improve the effectiveness of fuzzing in modern, stateful network systems such as 5G core components.

\subsection{Component Discovery analysis}

\begin{figure*}
    \centering
    \includegraphics[width=0.85\linewidth]{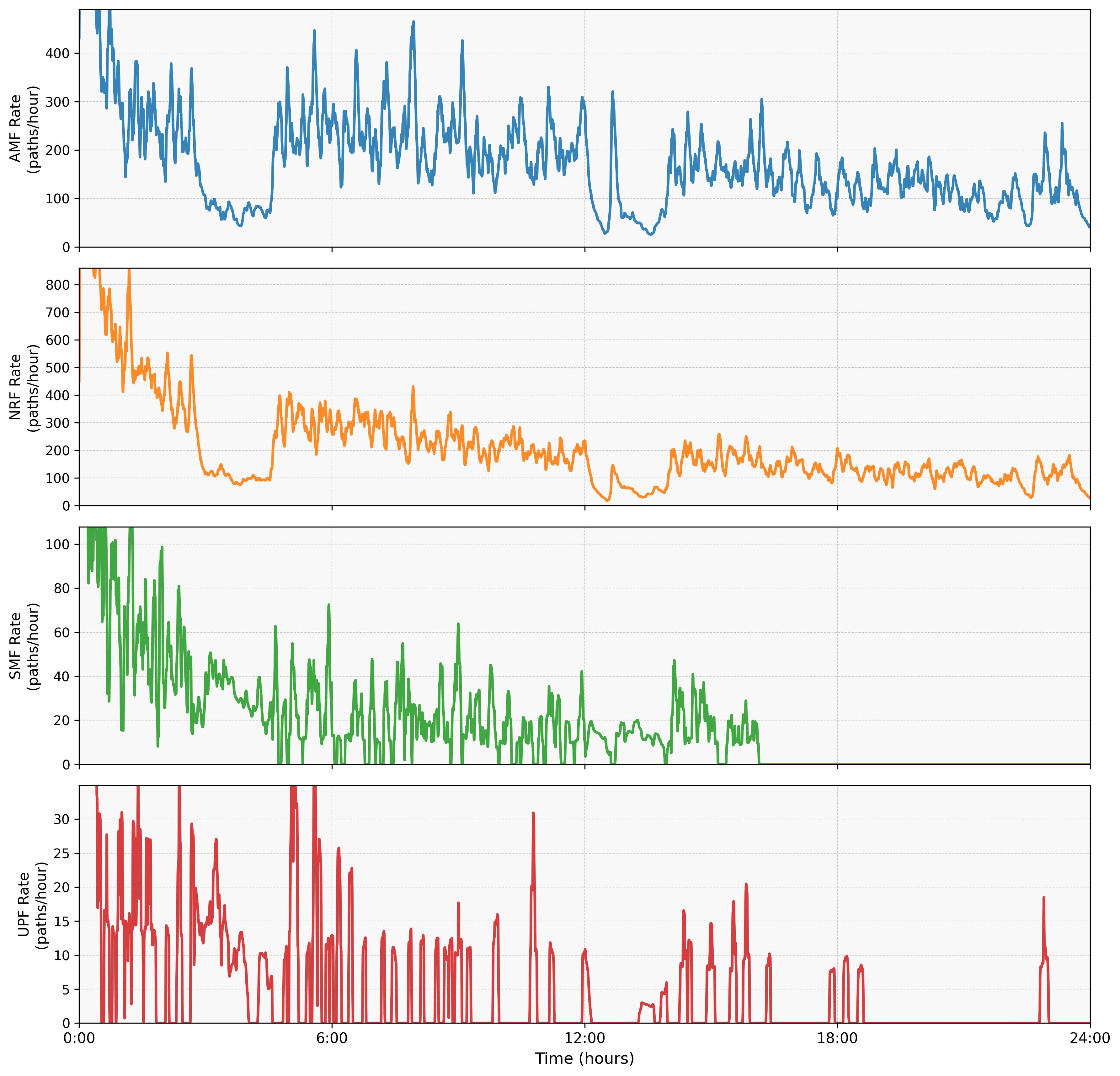}
    \caption{Interesting coverage discovery rate by each component in OAI}
    \label{fig:com_dis}
\end{figure*}

Fig. \ref{fig:com_dis} illustrates the interesting path discovery rate across four core network components (AMF, NRF, SMF, and UPF) during a 24-hour fuzzing campaign using MulCovFuzz. This visualization provides crucial insights into how the multi-component coverage approach affects exploration efficiency across different components.
All components exhibit high initial discovery rates during the first 3-4 hours, with AMF reaching peaks of approximately 450 paths/hour, NRF exceeding 800 paths/hour, SMF approaching 100 paths/hour, and UPF showing bursts of up to 35 paths/hour. This demonstrates the fuzzer's ability to rapidly explore basic functionality across the entire network stack during early stages.
The temporal correlation between discovery patterns is particularly noteworthy. Between hours 6-12, we observe synchronized fluctuations across components, suggesting that when MulCovFuzz discovers interesting paths in one component (particularly AMF), it often triggers cascading effects that lead to new path discoveries in other components. For example, the coverage spikes in AMF around hours 6, 8, and 11 are followed by corresponding increases in NRF and SMF coverage, demonstrating the interconnected nature of these components.
The significant drop in discovery rates observed across all components around hours 4-6 and 12-14 represents periods when the fuzzer transitions from exploring easily accessible paths to targeting more complex, state-dependent behaviors. After each drop, discovery rates recover but generally trend lower as the campaign progresses, indicating the increasing difficulty of finding unexplored paths.
The difference in discovery rates between components (AMF/NRF having substantially higher rates than SMF/UPF) reflects their architectural positions within the control plane. Components directly receiving fuzzed inputs (AMF) or serving critical control functions (NRF) experience more rapid path discovery, while components further downstream in the processing chain (SMF and particularly UPF) show more modest but still significant discovery patterns.
These results demonstrate that MulCovFuzz's multi-component approach successfully identifies complex interactions between components that would remain undiscovered using traditional single-component fuzzing approaches.

\subsection{Vulnerabilities analysis}

\begin{figure}
    \centering
    \includegraphics[width=1\linewidth]{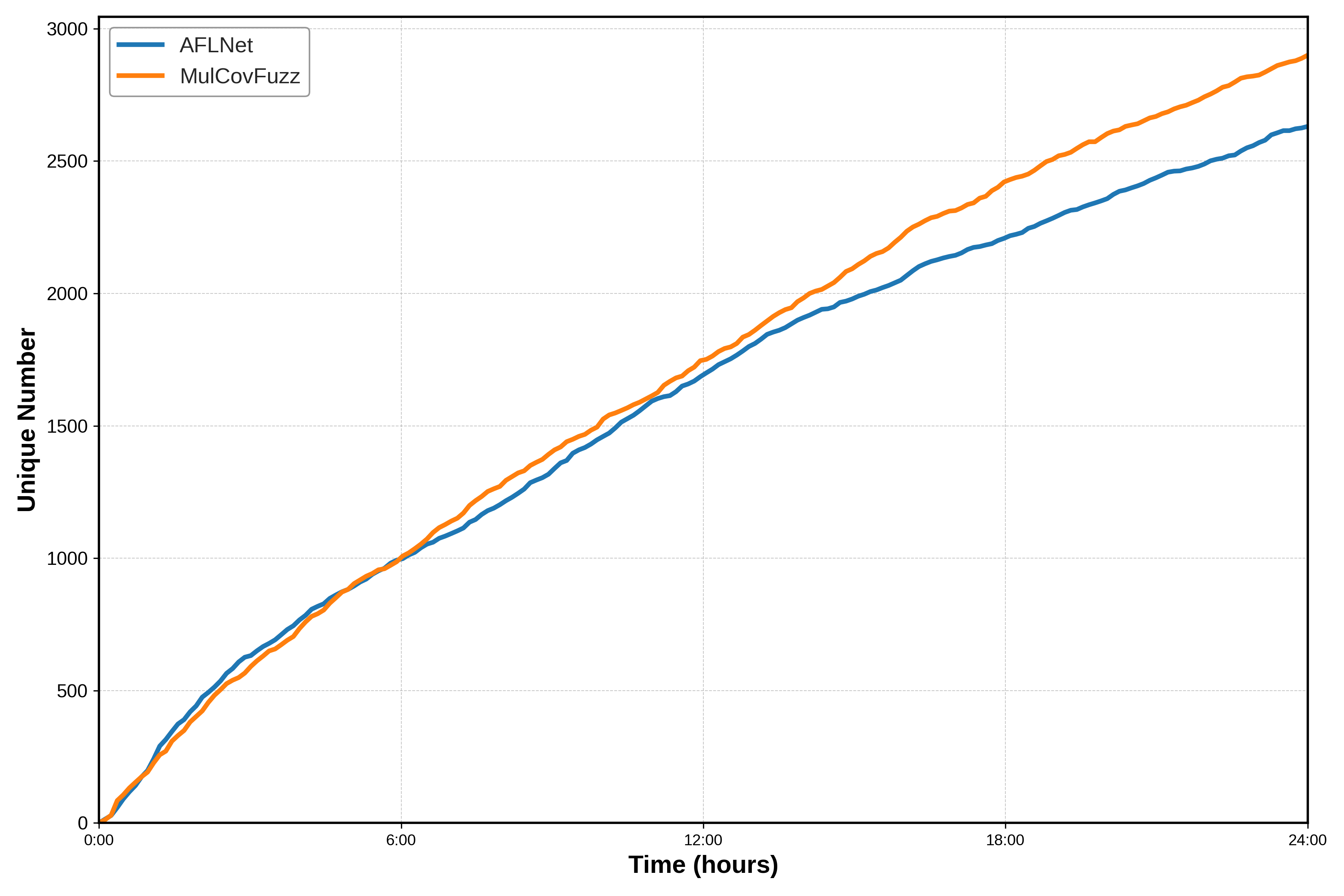}
    \caption{The unique crash number across 24h fuzzing in OAI-AMF}
    \label{fig:crash_num}
\end{figure}

In this section, we evaluate the vulnerability discovery capability of MulCovFuzz, a state-of-the-art stateful network protocol fuzzer. Fig. \ref{fig:crash_num} presents the average unique crash discovery trends over a 5 runs of 24 hour fuzzing campaign targeting the OAI-AMF implementation.

The results demonstrate that MulCovFuzz consistently discovers more unique crashes than AFLNet throughout the testing period. By the end of the 24-hour campaign, MulCovFuzz identified approximately 2900 unique crashes compared to AFLNet's 2500, representing a 16\% improvement in crash discovery efficiency. This significant advantage can be attributed to MulCovFuzz's multi-component coverage collection mechanism, which enables more thorough exploration of complex interaction scenarios between network components.
The crash discovery curve shows three distinct phases: (1) an initial rapid discovery phase (0-5 hours) where both fuzzers quickly identify obvious vulnerabilities, (2) a steady discovery phase (5-15 hours) with consistent but slower identification of new crashes, and (3) a late-stage discovery phase (15-24 hours) where MulCovFuzz continues to find new unique crashes while AFLNet's discovery rate diminishes.

In addition to the total number of crashes triggered by the fuzzer, it's important to note that multiple crashes may stem from the same underlying vulnerability but occur at different program execution paths. Consequently, the number of triggered crashes is typically higher than the actual number of distinct vulnerabilities. To address this, we leverage the program crash addresses reported by AddressSanitizer \cite{serebryany2012addresssanitizer} to manually analyze and deduplicate the program crashes, and collect statistics on the earliest discovery time for each distinct vulnerability type.

\begin{table}[]
\centering
\caption{Performance comparison of fuzzer in identifying vulnerabilities in OAI-AMF (in seconds)}
\label{table:vul_oai}
\resizebox{\linewidth}{!}{
\begin{tabular}{l l l l c l }
\hline
Id & Position              & Status  & Type  & \multicolumn{1}{l}{AFLNet} & MulCovFuzz \\ \hline
vul-01           & sctp\_server.cpp  &   CVE-2025-45983 (reserved)  & Null Pointer Dereference  & 20961                                              & 1937       \\
vul-02           & itti.cpp           & CVE requested  & Uncaught Exception  & $>$24h                                               & 58239      \\
vul-03           & 5gsMobileIdentity.cpp & CVE-2025-45982 (reserved) & Heap Buffer Overflow  & $>$24h                                               & 56530      \\ \hline
\end{tabular}
}
\end{table}

Table \ref{table:vul_oai} compares the effectiveness of different fuzzers in identifying vulnerabilities within the OAI-AMF implementation. The results highlight the superior capability of our multi-component coverage approach in uncovering diverse types of vulnerabilities within a limited time budget. MulCovFuzz discovered three zero-day vulnerabilities, two of which were not identified by any other fuzzer.

The analysis indicates that MulCovFuzz consistently outperforms AFLNet in detecting all identified vulnerabilities. For vul-01, a Null Pointer Dereference (NPD) in \textit{sctp\_server.cpp}, MulCovFuzz identified the vulnerability in 1,937 seconds, whereas AFLNet required 20,961 seconds. Moreover, vul-02, an uncaught exception in \textit{itti.cpp}, and vul-03, a Heap Buffer Overflow (HBO) in \textit{5gsMobileIdentity.cpp}, were both discovered by MulCovFuzz within the 24-hour testing window, while AFLNet failed to detect them entirely.

To further demonstrate the effectiveness of our approach, we present a case study of vulnerability vul-03, a HBO in the SUCI (Subscription Concealed Identifier) decoder function. The vul-03 occurs during the parsing of SUCI data within 5G Mobile Identity Information Elements in NAS Registration Request messages, representing a critical security flaw in the core mobility management functionality. MulCovFuzz successfully discovered this vulnerability at 56,530 seconds, while AFLNet failed to trigger it within the entire 24h testing period. The vulnerability's discovery required generating precisely crafted NAS Registration Request messages that establish proper context states across multiple 5g core network components to reach the vulnerable SUCI parsing logic, our multi-component coverage mechanism can effectively guide the fuzzer to explore.

\section{Discussion}

Our experimental results demonstrate that multi-component coverage feedback significantly enhances fuzzing effectiveness in complex 5G network environments. Several key insights emerge from this work that merit further discussion.

\subsection{Effectiveness of Multi-Component Coverage}
The superior performance of MulCovFuzz compared to traditional approaches can be attributed to its ability to capture interaction patterns across distributed components. By monitoring coverage across AMF, SMF, NRF, and UPF components simultaneously, MulCovFuzz successfully identified corner-case vulnerabilities that single-component fuzzers failed to detect. The synchronized coverage spikes observed across components (as shown in Fig. 5) confirm our hypothesis that test inputs triggering interesting behaviors often propagate through the entire network stack, affecting multiple components simultaneously.
However, this multi-component approach introduces significant implementation challenges. The need to instrument and monitor coverage across containerized, distributed components requires careful coordination to avoid excessive performance overhead. Our solution using lightweight collectors and asynchronous feedback mechanisms represents a practical compromise between coverage granularity and fuzzing throughput.

\subsection{Limitations and Challenges}
Despite its promising results, MulCovFuzz has several limitations that should be acknowledged. First, the current implementation focuses primarily on the N2 interface and AMF-initiated interactions. Extending this approach to cover other interfaces (N1, N3, N4) would require additional protocol parsers and state models. Second, the effectiveness of our approach depends on the quality of instrumentation across components, which can be challenging to maintain across different implementations and versions of 5G core networks.
Additionally, our current approach to coverage collection cannot effectively distinguish between "interesting" and "uninteresting" code paths without domain-specific knowledge. While we can observe coverage changes across components, determining which paths represent potential security vulnerabilities remains challenging. This is particularly true for logical vulnerabilities that don't necessarily manifest as crashes.

\subsection{Future Directions}
Several promising research directions emerge from this work. First, integrating domain-specific knowledge about protocol states and transitions could enhance the fuzzer's ability to target security-critical code paths. This could involve developing more sophisticated state models that capture cross-component dependencies and constraints.
Second, exploring adaptive weighting mechanisms for the scoring function could improve fuzzing efficiency. The current fixed weighting scheme ($\alpha_i$ values) could be replaced with a dynamic approach that adjusts weights based on observed coverage patterns during fuzzing.
Third, extending MulCovFuzz to support automated vulnerability classification and root cause analysis would significantly enhance its practical utility. This could involve integrating techniques from program analysis and symbolic execution to reason about the conditions leading to discovered vulnerabilities.
Finally, investigating the applicability of this multi-component coverage approach to other complex distributed systems beyond 5G networks would be valuable. The principles of multi-component coverage feedback could be beneficial for testing microservice architectures, IoT systems, and other distributed computing environments.

\section{Related work}

Prior studies have extensively explored protocol testing approaches for cellular networks, from early works on GSM \cite{van2014security} and UMTS \cite{meyer2004man} to more recent research on LTE security validation \cite{rupprecht2016putting, hussain2018lteinspector}. These efforts highlighted the challenges in systematically testing complex protocol interactions. For instance, Hussain et al. \cite{hussain2018lteinspector} developed LTEInspector, which combined symbolic model checking with cryptographic protocol verification to uncover security vulnerabilities in LTE control plane procedures. Their methodical approach of defining security properties and generating test cases based on protocol specifications laid important groundwork for testing next-generation cellular networks. Building upon these foundations, researchers have proposed various approaches to comprehensively test 5G core network implementations, particularly focusing on protocol conformance and robustness validation \cite{hussain20195greasoner, silveira2022tutorial}.

Traditional 5G fuzzing tools have evolved significantly but still face important limitations. Early works like Garbelini et al. \cite{garbelini2022towards} proposed fuzzing frameworks built on OpenAirInterface that supported testing across multiple protocol layers. However, their approach relied heavily on user-defined test cases and dynamically built state machines to detect improper transitions, requiring substantial manual effort. While tools like Berserker \cite{potnuru2021berserker} attempted mutation-based fuzzing of the RRC layer, they lacked adaptive strategies to guide the fuzzing process. Park et al. \cite{park2022doltest} and Kim et al. \cite{kim2019touching} developed testing frameworks based on careful analysis of 3GPP specifications, but this manual approach proved difficult to scale given the complexity of 5G protocols. A key limitation of traditional tools was their inability to automatically identify promising mutation targets, they either used fixed mutation strategies or required significant human intervention to generate effective test cases. Additionally, most tools operated without feedback mechanisms to guide the fuzzing process toward unexplored code paths, limiting their effectiveness in discovering deep protocol vulnerabilities. Pham et al. proposed the first grey-box fuzzing tool for network protocols, AFLNet \cite{pham2020aflnet}, in 2020. It extends traditional coverage-guided fuzzing by incorporating protocol state transitions to guide test case generation. By leveraging state transition information, AFLNet improves seed selection, enabling more efficient exploration of the protocol’s state space on the server under test. Building on this approach, the NCC Group utilized AFLNet to systematically fuzz a variety of protocols including GTP, PFCP, DIAMETER, and NGAP within the Open5GS 5G core network implementation.

\section{Conclusion}

In this paper, we presented MulCovFuzz, a novel coverage-based greybox fuzzing tool for 5G network. We introduced a multi-component coverage feedback mechanism that enables systematic monitoring of code coverage across multiple components of the 5G system architecture. This approach provides more testing effectiveness compared to traditional black-box testing methods. Then, we introduce a novel energy scoring function that prioritizes test cases based on their potential impact, thereby allocating more testing resources to those likely to exercise functional code related to inter-component interactions. Our experimental evaluation across three major open-source 5G implementation OpenAirInterface, demonstrates the effectiveness of our approach. The multi-component coverage feedback mechanism proved valuable in identifying previously unexplored code paths and potential vulnerabilities that might have been missed by conventional fuzzing approaches.

\begin{acks}
This research was conducted under the 6G Security Research and Development Project led by the Commonwealth Scientific and Industrial Research Organisation (CSIRO), with funding appropriated by the Australian Government Department of Home Affairs. The views expressed in this paper are those of the authors and do not necessarily reflect any Australian Government policy position. More information about the Project is available at \url{https://research.csiro.au/6gsecurity/}.
\end{acks}

\bibliographystyle{ACM-Reference-Format}
\bibliography{main}

\end{document}